\documentclass[prl,twocolumn,showpacs]{revtex4}

\usepackage{graphicx}
\usepackage{dcolumn}
\usepackage{bm}

\begin{document}

\preprint{ITP/UU-XXX}

\title{Theory for $p$-Wave Feshbach Molecules}

\author{K. B. Gubbels}
\email{K.Gubbels@phys.uu.nl}
\author{H. T. C. Stoof}

\affiliation{
Institute for Theoretical Physics, Utrecht University,\\
Leuvenlaan 4, 3584 CE Utrecht, The Netherlands}


\begin{abstract}
We determine the physical properties of \emph{p}-wave Feshbach
molecules in doubly spin-polarized $^{40}$K and find excellent
agreement with recent experiments. We show that these molecules
have a large probability $Z$ to be in the closed channel or bare
molecular state responsible for the Feshbach resonance. In the
superfluid state this allows for observation of Rabi oscillations
between the molecular and atomic components of the Bose-Einstein
condensed pairs, which contains a characteristic signature of the
quantum phase transition that occurs as a function of applied
magnetic field.
\end{abstract}


\maketitle

\emph{Introduction.} --- The continuous excitement in the field of
ultracold atoms is to a large extent caused by the ever increasing
experimental control over the creation and manipulation of
degenerate quantum gases. Arguably the most important example of
such control is the use of Feshbach resonances with which the
interaction between the atoms can be manipulated by simply varying
the magnetic field. The so-called \emph{s}-wave Feshbach resonance
occurs when atoms collide with zero orbital angular momentum
($\ell =0$) and is experimentally most easily accessible. It has
been widely used in the study of degenerate Fermi gases,
especially for the crossover between a Bardeen-Cooper-Schrieffer
(BCS) superfluid and a Bose-Einstein condensate (BEC) of diatomic
molecules \cite{Regal,Zwierlein,Thomas,Bartenstein,Salomon,Hulet}.

A novel challenge is to obtain superfluidity also with pairs in
nonzero orbital angular momentum states by using \emph{p}-wave
($\ell =1 $), or maybe even \emph{d}-wave ($\ell =2$) Feshbach
resonances. Up to now, \emph{p}-wave resonances have been seen in
Fermi gases of ${}^{40}$K \cite{Jin} and ${}^{6}$Li atoms
\cite{Zhang}. Most recently, Gaebler {\it et al}.\ have been able
to create and detect \emph{p}-wave Feshbach molecules in a gas of
${}^{40}$K with lifetimes on the order of milliseconds on both the
BEC and the BCS side of the resonance \cite{Gaebler}. The lifetime
on the BEC side, where the energy of the Feshbach molecule lies
below the open-channel continuum, is presumably limited by
collisional losses, whereas the lifetime on the BCS side is
determined by the decay of the molecular state into the
open-channel continuum. However, upon entering the superfluid
regime on the BCS side, the Bose-Einstein condensed pairs are
stabilized by the presence of a Fermi sea \cite{Romans}. Now that
their dominant decay mechanism is absent, the lifetime of the
pairs on the BCS side is expected to become significantly
enhanced, giving hope that \emph{p}-wave superfluidity is indeed
within reach.

There are many exciting aspects about \emph{p}-wave superfluidity.
First of all, a \emph{p}-wave superfluid has a more complex order
parameter than a \emph{s}-wave superfluid, due to the various
possible projections of its angular momentum ($m$=0,$\pm$1), which
can give rise to anisotropic superfluids. Second, a \emph{p}-wave
superfluid undergoes various classical and quantum phase
transitions as a function of temperature and magnetic field
\cite{Ho,Ohashi,Andreev,Yip,Iskin, Gurarie}. As an example, the
evolution from the BEC to the BCS side of the Feshbach resonance
at zero temperature contains a topological quantum phase
transition \cite{Read}, rather than the crossover known from the
\emph{s}-wave case. Third, the \emph{p}-wave resonances are
intrinsically narrow. As a result, the two-channel nature of the
Feshbach resonance becomes very prominent. This can be quantified
as follows.

The superfluid phase near a Feshbach resonance arises from the
Bose-Einstein condensation of pairs or so-called dressed molecules
\cite{Duine}, which are characterized by the linear superposition
$|\psi_{\rm dressed} \rangle = \sqrt{Z}|\psi_{\rm m}\rangle
+\sqrt{1-Z}|\psi_{\rm aa}\rangle,$ where $|\psi_{\rm m}\rangle$ is
the wave function of the bare molecular state in the closed
channel, and $|\psi_{\rm aa}\rangle$ describes the atom pairs in
the open channel. In the case of an \emph{s}-wave resonance, the
magnitude of the probability $Z$ can be estimated at resonance
(unitarity) by $Z \simeq 2 \sqrt{2 \Delta}/\pi \eta$
\cite{Romans}, where $\Delta$ is the universal BCS gap parameter
at unitarity and $\eta^2$ is a measure for the energy width of the
resonance. For the extremely broad \emph{s}-wave resonance of
${}^6$Li at 834 G, this leads at a Fermi energy of
$\varepsilon_{\rm F} = 380$ nK to $Z \simeq 4 \cdot 10^{-5}$. For
the \emph{s}-wave resonance of ${}^{40}$K at 202 G, we obtain at
the same Fermi energy $Z \simeq 5 \cdot 10^{-3}$, which is much
larger but still a rather small number. However, we will show
below that for a \emph{p}-wave resonance the value of $Z$ at
unitarity can become as large as $0.7$.

This shows that for typical \emph{p}-wave Feshbach resonances the
two-channel nature of the dressed molecules is very important.
Whereas \emph{p}-wave Feshbach molecules have recently been
carefully studied experimentally, theoretically the physical
properties of these molecules have not received much attention,
even though they are very interesting in their own right. In
particular, important quantities such as the probability $Z$, the
binding energy and the lifetime have not been determined yet in
terms of the experimentally known Feshbach parameters, i.e., the
background \emph{p}-wave scattering length $a_{\rm bg}$, the width
of the resonance $\Delta B$, and the magnetic moment difference
between the closed and the open channel $\Delta\mu$. In this
Letter, we show that given these parameters, all two-body
properties can be computed exactly and in excellent agreement with
experiment. Moreover, we perform a two-channel many-body
calculation to compute the frequency of coherent Rabi or Josephson
oscillations between the closed and open channel components of the
Bose-Einstein condensed pairs.

\emph{Two-body physics.} --- The two-body problem near a
\emph{p}-wave Feshbach resonance can be treated along the same
lines as the corresponding problem for the \emph{s}-wave resonance
\cite{Duine}. We consider two fermionic particles that are in the
same hyperfine spin state and that interact with each other
through a coupling with a closed molecular channel and through a
\emph{p}-wave background interaction in the open channel, both
with angular momentum projection $m$. The scattering due to the
background interaction generates a ladder sum, which can be
conveniently incorporated into the atom-molecule coupling
\cite{Duine}, and gives rise to a dressed coupling $g_{m}({\bf
k})=\langle \psi_{\rm m}|V_{\rm ma}|\psi^{(+)}_{\bf
k}\rangle/\sqrt{2}= g_{m} k_{m}/[1+i (k a_{\rm bg})^3]$, where
$|\psi^{(+)}_{\bf k}\rangle$ is the scattering state with momentum
$\bf{k}$ in the open channel, $V_{\rm ma}$ the coupling potential
between open and closed channels, and $a_{\rm bg}$ is directly
related to the effective hard-core radius $a_{\rm hc}$ of the
background potential by $a_{\rm bg}=a_{\rm hc}/3^{1/3}$. The
subscript $m$ refers to the projection of the angular momentum
$m=0,\pm 1$, such that $k_0=k_z$ and $k_{\pm 1}=\mp(k_x \pm i
k_y)/\sqrt{2}$, while the coupling constant $g_m$ characterizes
the strength of atom-molecule interactions. Note that the dressed
or renormalized coupling $g_{m}({\bf k})$ is exact at low
energies.

With this dressed atom-molecule coupling we can calculate the
two-body molecular self-energy analytically,
\begin{eqnarray}
\hbar\Sigma_{m}(z)&=&\frac{2}{V}\sum_{\bf k}|g_{m}({\bf
k})|^2\left(\frac{1}{z-2\varepsilon_{\bf k}}+
{\frac{1}{2\varepsilon_{\bf k}}}\right) \nonumber\\
&=& \frac{m_{\rm a} g_{m}^2}{18 \pi \hbar^2|a_{\rm
bg}|^3}\frac{-\zeta(2+\sqrt{-\zeta})}{1+2\sqrt{-\zeta}- \zeta
(2+\sqrt{-\zeta})},\label{selfen}
\end{eqnarray}
where $\varepsilon_{\bf k}=\hbar^2 k^2/2m_{\rm a}$ is the kinetic
energy with $m_{\rm a}$ the mass of a single atom. The
dimensionless energy $\zeta$ is given by $\zeta=m_{\rm a} a_{\rm
bg}^2 z/\hbar^2$. To obtain this result we have renormalized the
bare self-energy $\hbar\Sigma^{\rm B}_m(z)$ by subtracting
$\hbar\Sigma^{\rm B}_m(0)$. This is convenient because in the
equation for the bound state energy of the dressed molecule
\cite{Duine}, namely $E-\delta_m^{\rm B} = \hbar\Sigma^{\rm
B}_m(E)$, we can subtract $\hbar\Sigma^{\rm B}_m(0)$ on both
sides, which also renormalizes the bare detuning $\delta_{m}^{\rm
B}$ to the renormalized detuning $\delta_{m}$ and leads to
$E-\delta_{m} = \hbar\Sigma_m(E)$. The two-body resonance now
indeed occurs when the bound-state energy reaches the atomic
continuum, i.e., at $E=0$ and $\delta_{m} =0$. Therefore, the
renormalized detuning is the experimentally relevant quantity and
given by $\delta_{m}=\Delta \mu (B-B_{m})$, with $B$ the applied
magnetic field and $B_{m}$ the location of the Feshbach resonance.
Note that in general the Feshbach resonances with different values
of $m$ need not be located at the same magnetic field \cite{Bohn}.
This is because only the projection of the total angular momentum
is conserved in the collision of two alkali atoms. Also note that
by using the dressed atom-molecule coupling, the right-hand side
of Eq. (\ref{selfen}) is finite, so no arbitrary, undetermined
cutoff needs to be introduced. For small energy we obtain that
$\hbar\Sigma_{m}(z) \simeq (g_m^2 m_{\rm a}/\hbar^2)[-m_{\rm a}
z/9\pi\hbar^2|a_{\rm bg}| - (-m_{\rm a} z/\hbar^2)^{3/2}/6\pi]$.
The `universal' coefficient of $(-m_{\rm a} z/\hbar^2)^{3/2}$ is
in agreement with Refs.\ \cite{Gurarie,Chevy}.

\begin{figure}
\includegraphics[width=0.93\columnwidth]{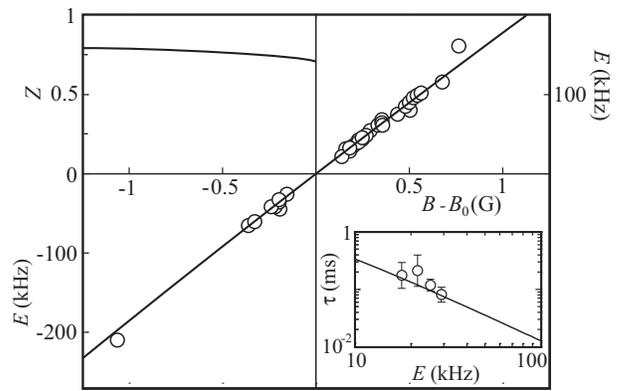}
\caption{\label{Twobody} Two-body physics near the $m=0$ $p$-wave
resonance of ${}^{40}$K located at $B_0= 198$ G. The upper left
panel shows the probability $Z$ to be in the closed Feshbach
channel as a function of the magnetic field $B$. In the lower left
and the upper right panel the black line shows the binding energy
$E$ as a function of magnetic field. For positive detuning the
molecules are not stable, resulting in a finite lifetime $\tau$,
shown in the lower right panel. The open circles are data from the
experiments of Gaebler {\it et al}.}
\end{figure}

The binding energy of \emph{p}-wave Feshbach molecules in a gas of
spin-polarized ${}^{40}$K atoms near the $m=0$ Feshbach resonance
at 198 G has been accurately measured as a function of magnetic
field by Gaebler {\it et al}.\ \cite{Gaebler}, where they find an
almost perfectly linear behavior. From the slope, which is given
by $Z \Delta \mu$, we obtain $\Delta \mu=0.175 \mu_{\rm B}$ with
$\mu_{\rm B}$ the Bohr magneton. However, for small detuning they
report tiny deviations in this linear behavior. In Fig. 1 we have
calculated the binding energy with the use of Eq. (\ref{selfen}),
where we extract the relevant Feshbach parameters $a_{\rm bg}$ and
$g_0$ from the collisional phase shift of the considered
\emph{p}-wave resonance obtained in Ref.\ \cite{Bohn}. This
results in $a_{\rm bg} = 98 a_0$ and $g_0 = 33 a_0^{1/2}
\hbar^2/m_{\rm a} $, with $a_0$ the Bohr radius. We see that the
agreement with our two-body calculation is excellent. The
nonlinearities arise from the slight variation of $Z$ with
magnetic field.

Having obtained the self-energy and the equation for the binding
energy of the molecules, we can also determine the two-body wave
function renormalization factor $Z$, given by $Z=[1-\partial \hbar
\Sigma_{m}(E)/\partial E]^{-1}$ with $E$ the solution of the
equation for the binding energy. The behavior of $Z$ as a function
of magnetic field is shown in Fig. 1. We indeed observe that for
the considered \emph{p}-wave resonance $Z$ is very large at
unitarity, namely $Z \simeq 0.7$, which proves the two-channel
nature of the \emph{p}-wave molecules near this resonance. Since
the difference in magnetic moment between a dressed molecule and
two atoms is $Z \Delta \mu$, we see that this difference does not
go to zero at a \emph{p}-wave resonance. As a result, the binding
energy of the dressed molecule vanishes only as $(B-B_{m})$ and
not as $(B-B_{m})^2$, which is the well-known behavior from the
$s$-wave case \cite{Yip,Chevy}.

In the two-body case, there are only stable molecules for negative
detuning and strictly speaking $Z$ ceases to have meaning for
positive detuning. In the many-body case the Feshbach molecules
are also stable at the BCS side, due to the presence of a Fermi
sea, preventing decay into the atomic continuum \cite{Romans}. It
is then possible to determine $Z$ at the BCS side using many-body
techniques, which has been done for the \emph{s}-wave case both
theoretically \cite{Romans} and experimentally \cite{Hulet}. In
the two-body case the \emph{p}-wave molecules are not stabilized
by a Fermi sea and the decay rate $\Gamma_{\rm m}$ is readily
calculated from the self-energy by $\Gamma_{\rm m} = -2 {\rm
Im}[\Sigma_{m}(E+i0)]$. The lifetime is thus $\tau_{\rm
m}=1/\Gamma_{\rm m}$. The calculated lifetimes are shown in Fig.
1, leading to very good agreement with the experimentally observed
lifetimes without any adjustable parameters.

\emph{Internal Josephson oscillations.}---  Having obtained the
analytic solution to the two-body \emph{p}-wave Feshbach problem,
the next step is to incorporate this two-body physics into the
many-body physics. To this end we use the following effective
grand-canonical Hamiltonian with chemical potential $\mu_{\rm a}$
for the atoms and $\mu_{\rm m}$ for the molecules,
\begin{eqnarray}
H &=& \sum_{\bf k}(\varepsilon_{\bf k}-\mu_{\rm
a})a^{\dagger}_{\bf k}a^{\phantom \dagger}_{\bf k}+\sum_{\bf
k}\left(\frac{\varepsilon_{\bf k}}{2}+\delta_m-\mu_{\rm
m}\right)b^{\dagger}_{m,{\bf
k}}b^{\phantom \dagger}_{m,{\bf k}} \nonumber\\
&&+\frac{1}{\sqrt{V}}\sum_{\bf p,k}\left(g_{m}({\bf k})
b^{\dagger}_{m,{\bf p}}a^{\phantom \dagger}_{{\bf p}/2+{\bf
k}}a^{\phantom \dagger}_{{\bf p}/2-{\bf k}}+{\rm h.c.}
\right),\label{Hamil}
\end{eqnarray}
where $a^{\dagger}_{\bf k}$ creates an atom with momentum ${\bf
k}$, $b^{\dagger}_{m,{\bf k}}$ creates a bare \emph{p}-wave
molecule with angular momentum projection $m$ and momentum ${\bf
k}$, and $V$ is the volume of the gas. This Hamiltonian is valid
when different values of $m$ lead to resonances at well-separated
magnetic fields $B_m$, as is experimentally the case for the $m=0$
resonance of ${}^{40}$K. If this is not true, then on the
right-hand side of Eq. (\ref{Hamil}) we also have to perform a sum
over $m$. Furthermore, in equilibrium we have that $\mu_{\rm
m}=2\mu_{\rm a}$, but for reasons that become clear shortly, we
enforce this relation only at the end of the calculations. For
\emph{p}-wave resonances the total scattering length satisfies
$a^3 = a^3_{\rm bg}-m_{\rm a} g_m^2/6 \pi \hbar^2 \delta_m$. If we
parametrize this in analogy with the $s$-wave case as $a^3 =
a^3_{\rm bg}[1-\Delta B_m/(B-B_m)]$, we see that the atom-molecule
coupling obeys $g_m = \hbar (6\pi \Delta\mu \Delta B_m a^3_{\rm
bg}/m_{\rm a})^{1/2}$, with $\Delta B_m$ the magnetic field width
of the Feshbach resonance. Near resonance the background
interaction may be safely neglected in the Hamiltonian, except for
its effect on the atom-molecule coupling $g_m({\bf k})$. The
reason for the latter is that, although the effective Hamiltonian
above is designed to be accurate for low energies, we still need
high-energy states to properly describe the two-body physics near
a Feshbach resonance. As we have seen, the effect of these states
is to renormalize the atom-molecule coupling and therefore it is
essential to use the dressed coupling $g_m({\bf k})$ in the
effective Hamiltonian \cite{Romans}.

In order to discuss internal Josephson oscillations \cite{Legett,
Galitski}, we turn to the functional-integral formalism
\cite{Duine} in which the creation and annihilation operators
become complex fields that not only depend on momentum, but also
on time $t$. A sketch of the derivation goes as follows. First, we
rewrite the atomic and molecular fields as $a_{\bf k}(t)
\rightarrow a_{\bf k}(t) e^{i \theta_{\rm a}(t)}$, $b_{m,{\bf
k}}(t) \rightarrow b_{m,{\bf k}}(t) e^{i \theta_{\rm m}(t)}$ to
explicitly consider the atomic and molecular phase fluctuations
$\theta_{\rm a}(t)$ and $\theta_{\rm m}(t)$, respectively.
Substituting these expressions in the action, it changes in the
following way: $g_m({\bf k}) \rightarrow g_m({\bf k})e^{i [2
\theta_{\rm a}(t)-\theta_{\rm m}(t)]}$ and $\mu_{j}\rightarrow
\mu_{j} - \hbar \dot{\theta}_{j}(t)$, where the last substitution
is due to the time-derivatives in the action and the index $j$
runs over two possible subscripts, namely atomic and molecular.
Next, we separate the action into the part $S_{\rm ph}$ containing
the phase fluctuations, and the part $S$ that doesn't contain
these fluctuations. This allows us to do perturbation theory in
the phase fluctuations by expanding the exponential containing
$S_{\rm ph}$. Expanding in the fluctuations up to second order, we
see that the coefficients in front of the phase fluctuations get
averaged over the original action $S$. For example, the term
linear in $\dot{\theta}_{\rm m}(t)$ becomes $\int
dt~\dot{\theta}_{\rm m}(t) \sum_{\bf k} \langle b^*_{m,{\bf k}}(t)
b_{m,{\bf k}}(t) \rangle$, where the average $\sum_{\bf k} \langle
b^*_{m,{\bf k}}(t) b_{m,{\bf k}}(t) \rangle$ equals the total
number of bare molecules and is conveniently expressed as
$-\partial \Omega(\mu_{\rm a},\mu_{\rm m})/\partial \mu_{\rm m}$
with $\Omega(\mu_{\rm a},\mu_{\rm m})$ the thermodynamic potential
of the gas. By re-exponentiating all terms up to quadratic order,
we obtain the effective action for the phase fluctuations,
\begin{eqnarray}
S^{\rm eff}[\theta_j]&=&\int dt \Bigg\{ \sum_{j}
\hbar\dot{\theta}_j\frac{\partial\Omega}{\partial \mu_j}-
\sum_{j,j'}
\frac{\hbar^2}{2}\dot{\theta}_j\dot{\theta}_{j'}\frac{\partial^2\Omega}{\partial
\mu_j\partial \mu_{j'}}  \nonumber \\
&&   + J \cos(2\theta_{\rm a}-\theta_{\rm m})
\Bigg\},\label{effact}
\end{eqnarray}
where the Josephson coupling is given by the expression $J=
2\sum_{\bf k} g_{m}({\bf k}) \langle b^*_{\bf 0}(t) a_{\bf k}(t)
a_{-{\bf k}}(t) \rangle/\sqrt{V}$.

\begin{figure}[t]
\includegraphics[width=0.93\columnwidth]{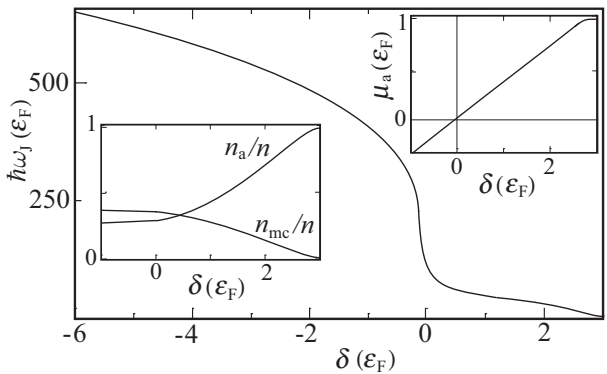}
\caption{\label{densprof} Josephson frequency $\omega_{\rm J}$ as
a function of the detuning $\delta$. Energies are expressed in
terms of the Fermi energy $\varepsilon_{\rm F}=\hbar^2(6 \pi^2
n){}^{2/3}/2m_{\rm a }$, with $n=2n_{\rm mc}+n_{\rm a}$ the total
density of atoms, which is kept fixed during the calculation at
$n=3 \times 10^{18}$ m$^{-3}$, so that $\varepsilon_{\rm F}/ \hbar
\simeq 2 \pi \times 4$ kHz. In the left inset, the number of bare
molecules $n_{\rm mc}$ and open-channel atoms $n_{\rm a}$ are
given as a function of the detuning. In the right inset we have
the atomic chemical potential vs. detuning.}
\end{figure}

In deriving the quadratic action for the phase fluctuations we
made one approximation, namely a gradient expansion of the
fluctuations that takes into account static number correlations.
To calculate the coefficients in Eq. (\ref{effact}) we make
another approximation, namely mean-field theory. The mean-field
thermodynamic potential for a Fermi gas near a \emph{p}-wave
resonance at zero temperature is the direct generalization of the
\emph{s}-wave case, giving
\begin{eqnarray}
\Omega &=& \frac{1}{2}\sum_{\bf k}\left\{\varepsilon_{\bf
k}-\mu_{\rm a}-\hbar \omega_{\bf k}+\frac{2 |g_m({\bf k})|^2
n_{\rm mc
}}{\varepsilon_{\bf k}}\right\} \nonumber \\
&&+(\delta_m-\mu_{\rm m})n_{\rm mc}V,
\end{eqnarray}
with $\hbar \omega_{\bf k}=\sqrt{(\varepsilon_{\bf k}-\mu_{\rm
a})^2+4 |g_m({\bf k})|^2 n_{\rm mc}}$ and $n_{\rm mc}$ the bare
molecular condensate density. The gap equation follows from
differentiation with respect to $n_{\rm mc}$, yielding
\begin{equation}
\delta_m-\mu_{\rm m} =\frac{1}{V}\sum_{\bf k}|g_m({\bf k})|^2
\left(\frac{1}{\hbar \omega_{\bf k}}-\frac{1}{\varepsilon_{\bf k}}
\right),
\end{equation}
which can be seen as the many-body generalization of the two-body
bound state equation $E-\delta_{m} = \hbar\Sigma_{m}(E)$, with
$E=\mu_{\rm m}$. Within mean-field theory, the coefficients in Eq.
(\ref{effact}) are readily determined. For example,
$-\partial\Omega/ \partial \mu_m=n_{\rm mc}V$, where $n_{\rm mc}$
is obtained from the gap equation. Also, we have that
$-\partial\Omega/ \partial \mu_a=n_{\rm a}V=\sum_{\bf
k}[1-(\varepsilon_{\bf k}-\mu_{\rm a })/\hbar\omega_{\bf k }]/2$,
with $n_{\rm a}$ the atomic density. Even the second derivatives
are given by simple analytic expressions in terms of momentum
integrals. Finally, the Josephson coupling is given by $J=2 n_{\rm
mc}\sum_{\bf k}|g_m({\bf k})|^2/\hbar\omega_{\bf k}$. These
expressions all depend quantitatively on $a_{\rm bg}$, showing
that the background interaction is important.

From the effective action we can then derive the equations of
motion by writing down the Euler-Lagrange equations. The remaining
second-order differential equations can be solved analytically,
and gives rise to two types of modes. One has frequency zero and
is recognized as the Goldstone mode due to the spontaneous broken
$U(1)$ symmetry associated with the conservation of the total
number of atoms in the gas. The other mode has a nonzero frequency
$\omega_{\rm J}$ and is recognized as the collective mode
corresponding to internal Josephson or Rabi oscillations between
the molecular and the atomic components of the Bose-Einstein
condensed pairs. In Fig. 2, we show the Josephson oscillation
frequency, the atomic chemical potential, the atomic density and
the bare molecular density as a function of the detuning with the
parameters for the $m=0$ Feshbach resonance of ${}^{40}$K. Far
below resonance, not visible on the scale of the figure, we find
that $\omega_{\rm J} \simeq \delta$. However, near resonance,
where the detuning is no longer the dominant energy scale, we
observe completely different behavior. In particular, on approach
of the quantum critical point at $\mu_{\rm a}=0$ \cite{Read}, we
find that the slope of the Josephson frequency develops a
logarithmic divergency. Since internal Josephson oscillations can
be accessed experimentally by performing a Ramsey-type experiment
\cite{Donley}, this characteristic feature in the Josephson
frequency is an observable signature for the occurrence of the
quantum phase transition. Although the Josephson oscillations are
damped by pair breaking, we expect this effect to be suppressed
since the single-atom continuum is Pauli-blocked by the Fermi sea
and the Bose-Einstein condensed pairs are protected by the
(mostly) gapped atomic spectrum. Also note that the collective
mode of Josephson oscillations is intrinsically connected to the
two-channel nature of $p$-wave resonances and is absent in a
single-channel treatment of the interactions. Finally, we hope
that our calculation will inspire further research towards
$p$-wave superfluidity.

\emph{Acknowledgements.}
--- We thank Jayson Stewart for kindly providing us with the experimental data. This work
is supported by the Stichting voor Fundamenteel Onderzoek der
Materie (FOM) and the Nederlandse Organisatie voor Wetenschaplijk
Onderzoek (NWO).

\end{document}